\documentclass[aps,pre,onecolumn,epsf,graphicx,a4]{revtex4}
\usepackage[pdftex]{graphicx,color}    
\usepackage{amsfonts}
\usepackage{epsfig}
\usepackage{amsmath}
\usepackage{amssymb}

\def\pd2x{{\partial^2 \over \partial x^2}}

\usepackage{epsfig,latexsym}

\begin{document}

\title{\bf\noindent Searching for Nodes in Random Graphs}

\author{David Lancaster}

\affiliation{
Department of Computing and Mathematics, University of Plymouth,
Drake Circus, Plymouth PL4 8AA, UK.
}
\date{\today}
\begin{abstract}
We consider the problem of searching for a node on a labelled
random graph according to a greedy algorithm that selects
a route to the desired node using metric
information on the graph. 
Motivated by peer-to-peer networks 
two types of random graph are proposed with
properties particularly amenable to this kind of algorithm.
We derive equations for the probability that the search is successful
and also study the number of hops required
to find both numerical and analytic evidence of a transition as the
number of links is varied.
\end{abstract}

\maketitle
\vspace{.2cm} 
\pagenumbering{arabic}

\section{Introduction}

More than a decade ago, small world graphs were proposed
\cite{SmallWorld} to model networks found in nature
and this subsequently led to an explosion of interest in
the topic.
The shortest path between two nodes in a small world graph is
typically no more than $\sim\log N$ where $N$ is the size of the graph.
However, although graphs with these short paths can be constructed, 
the way in which the path or route to a specified destination node
is discovered is another matter.
Typical approaches are based on recursively
flooding enquires to all neighbours on the graph and this kind of
global propagation is the basis of the shortest path algorithms such
as Dijkstra or Bellman-Ford. These approaches suffer 
from a consequent lack of scalability
and indeed routing on the internet uses a hierarchical approach
to counteract the problem.
Modern peer-to-peer protocols pursue a different approach
to avoid the scaling difficulty by having a particular
network structure that allows a small number of select queries to 
efficiently find the required destination.

This paper considers the problem of how to find the shortest path to a given
labelled node on a random graph without relying on global propagation.
Our approach to the problem relies on two essential ingredients:
a simple rule acting on local information 
at each node and an appropriately chosen random graph structure. 
The complication of the peer-to-peer systems which motivate the study 
are discarded to choose these ingredients in the simplest way 
that exposes the problem.
Firstly we assume that the routing rule  greedily
attempts to get as close as possible to the destination
at each step. For this to work, the graphs we consider must 
have a small world like structure, that is, they have
short characteristic path length but tend to cluster nodes with
similar labels together.
Moreover, in order to simplify the analysis
by allowing local information to uniquely specify which link to 
take at intermediate steps, we require that the graphs satisfy
a strong version of the triangle inequality  determining,
not merely bounding, the third side of a triangle in terms of the
other two. These requirements are not satisfied by the
random relinking construction of Watts and Strogatz \cite{SmallWorld}, and 
instead  our random graphs are based on a modification the traditional 
construction of
Erd\H{o}s and R\'enyi  \cite{RandomGraph} that selects links
to delete from the fully connected graph according to
a probability depending on their metric weight.
To encourage clustering of nearby nodes  we
only consider probability distributions that favour
short links.

Having mentioned peer-to-peer systems as a motivation, it
may be helpful to briefly review aspects of modern distributed
hash tables to clarify the relationship to our work.
These systems store data in a distributed setting by associating the
data with an integer key (the hash), and place the data on a computer node
with integer nodeID close in value to the key. 
The essential function that these systems provide is to
allow any node to efficiently retrieve the data and this
becomes the ability to locate a particular node according to its nodeID.
These nodes abide on a computer network that allows 
packets to be sent to any node given its computer address. However
the address and the nodeID are distinct entities and
nodes only store a table of addresses corresponding to a small
fraction of the total number of nodes. A node, and the data
stored on it, is located by a  series of queries to other
nodes that return the addresses from their own tables that
are closest to the desired node. The queries are
first sent to the nodes in the local table that are closest
to the nodeID being sought, and then recursively to the 
addresses returned from the queries. This process can involve
multiple queries at each stage and in the event that a
query fails to result in a closer node address, other 
queries are attempted. Through these mechanisms
nodes are located successfully  and efficiently 
with overwhelming probability.

Two well studied peer-to-peer systems are Chord \cite{Chord}
and Kademlia \cite{Kademlia} and the metrics they use to
determine the closeness of nodeID's, distance around a circle and
XOR respectively, are exactly those used as the metrics
on the random graphs in this paper. Nodes on peer-to-peer systems
use the metric to organise their table of nodeID addresses 
to contain more addresses of nearby nodeID's than distant nodeID's.
The random graphs in our work can 
be thought of as being determined by the connectivity implied by these
tables, and will have clustering as a consequence.
The degree of a node in the random graph should therefore
be determined by the size of the table in the peer-to-peer system.
Although this table size appears to be configurable, it 
must in fact grow according to the log of the maximum number of nodeID's 
and the degree of a node in the random graph also grows 
as $\log N$.
However, peer-to-peer systems are dynamic, with nodes continually entering
and leaving the system so the nodeID's are sparse and
the local tables themselves are
frequently updated. This dynamic aspect is not part of
our work where we take the nodes to be labelled contiguously
and consider static graphs with randomly generated connectivity.
The series of queries that occur in a node lookup in a peer-to-peer
system is replaced in our work by the stepwise deterministic construction
of a path along neighbouring links of the random graph towards the
desired node. 
Crucially, in our work we use a greedy algorithm and at each step discard
all routes that are not the best, so the
path can arrive at a dead end where no neighbour is closer
to the final destination. In this case the search has failed.

In summary, the context within which we study this problem
is that of a simple greedy algorithm and a rather
complex random graph structure. While peer-to-peer systems
motivate this context, the algorithm used is different and
their overlay networks are certainly not random graphs.
Moreover, in our analysis we are concerned with behaviour
in the large graph limit. 
We consider the probability of  successful search and
in particular its asymptotic value as the sought node 
becomes further and further away. We also measure statistics 
for the number of steps in a successful path. 
The most interesting observation is that there appears 
to be a transition in the probability of success as 
the average connectivity is varied. This is quite distinct 
from the percolation transition as all the random graphs we
consider will consist of a single connected component.

Subsequent sections will discuss the construction of the
random graphs in detail and study some of their properties. The greedy
routing algorithm is then introduced and basic 
equations used to analyse  the algorithm are
derived. The analysis itself constitutes the main
section and considerable attention is paid to one model
that can be solved exactly. However, although this solution provides
clues that are used for approximate analysis, we rely heavily
on numerical solution of the equations and also check our
results against simulation of the whole system.


\section{Graphs}

We shall consider graphs constructed in a manner
similar to traditional Erd\H{o}s and R\'enyi random
graphs\cite{RandomGraph} (as opposed to the configuration
approach of Molloy and Reed\cite{ReedMolloy}), by
diluting the links of a fully connected metric graph. First 
imagine a fully connected graph on $N$ nodes
labelled $0,1,2\dots N-1$ in which the link
between nodes labelled $a$ and $b$ 
has length $d_{ab}$ according to the graphs's metric.
Links will be selected to appear in the random graph
according to a length dependent probability distribution. 
In order that the graph be uniform, in the sense that 
all nodes are statistically equivalent, the set of link lengths 
emerging from any node in the fully connected graph 
must be the same, and for the graphs we consider will take the discrete values
$1,2,3\dots N-1$. 

We shall imagine that each node knows the
labels of the neighbouring nodes to which it is directly connected,
and nothing else. Specifically it has no knowledge of 
who its neighbours are connected to. 
This avoids undue complexity such as routing updates
and complex forwarding tables at each node and 
radically simplifies the dynamic nature of 
peer-to-peer networks.
In order that this local information is sufficient to
allow the greedy algorithm to determine which 
link to use to get closest to the eventual goal,
we consider metrics having the property that the
length of the third side of a triangle is completely
determined by the lengths of the other two sides.
Two metrics with the necessary properties
are the circle and XOR metrics.

\begin{figure}[htbp]
  \centering
  \begin{minipage}[b]{5 cm}
\epsfig{file=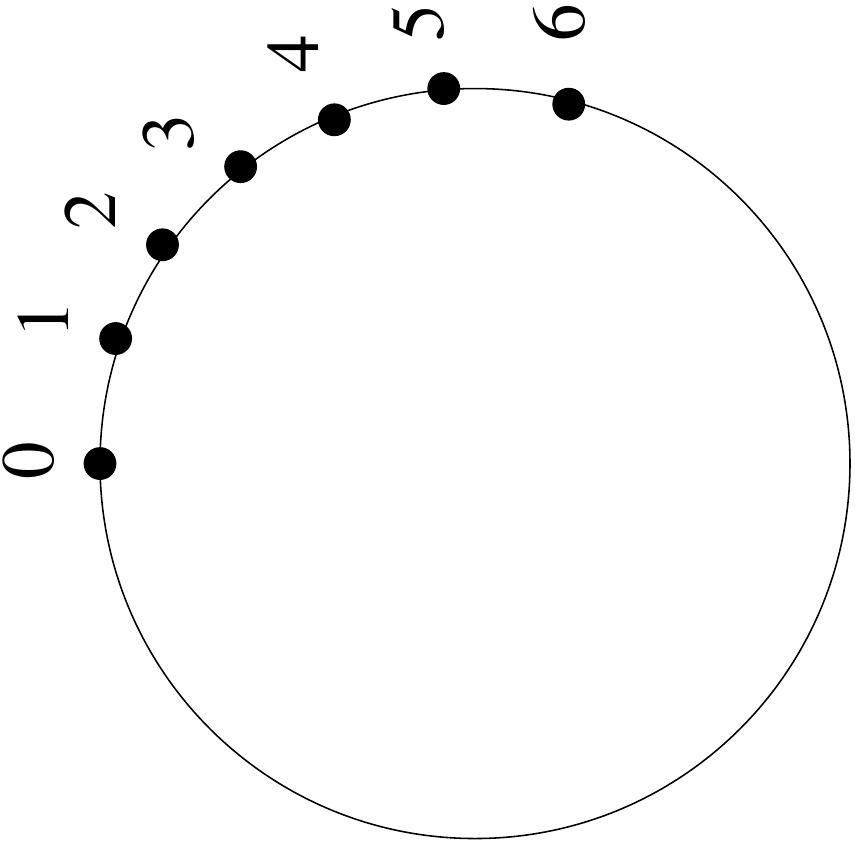,width=4cm,angle=270}
\caption{For the circle metric
the distance between nodes is measured clockwise round the circle.}
\label{fcircle}
  \end{minipage}
\qquad
  \begin{minipage}[b]{5 cm}
\epsfig{file=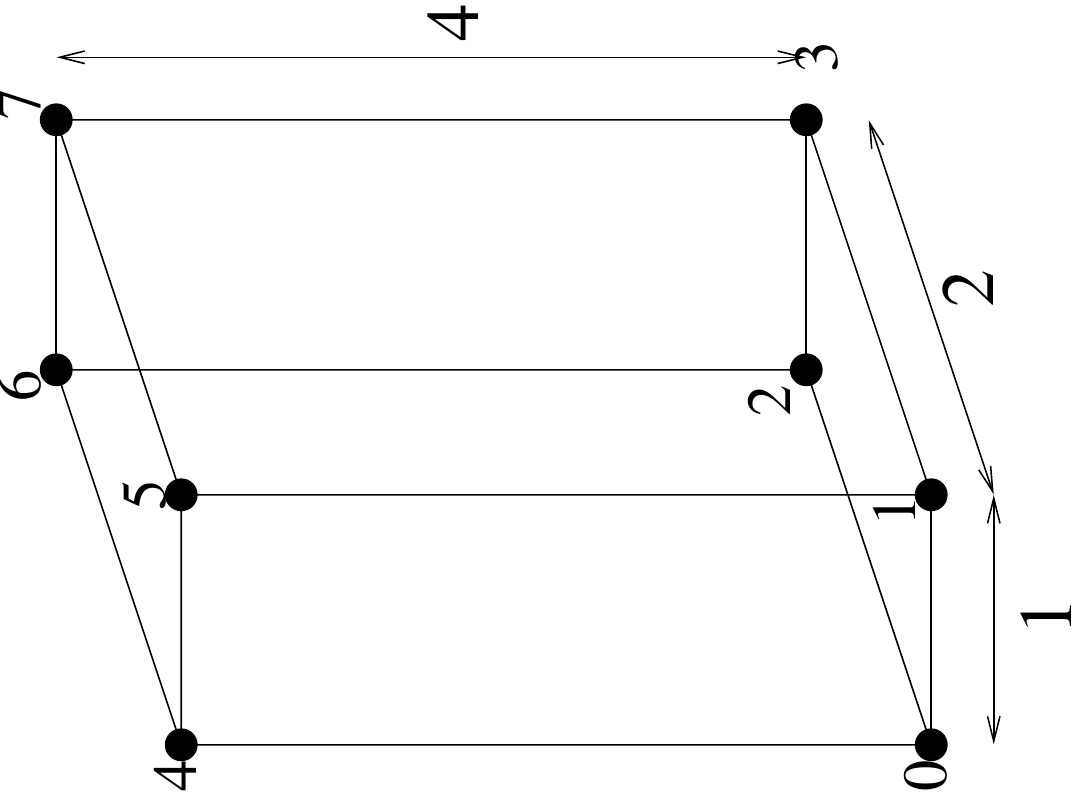,width=4cm,angle=270}
\caption{The Manhattan distances between nodes with the
XOR metric for 8 points.}
\label{fXOR}
  \end{minipage}
\end{figure}

\bigskip
\noindent{\bf Circle metric.} 
The length of a link from node labelled $a$ to $b$ is:
\begin{equation}
d_{ab} = (b-a) mod N
\end{equation}
This is simply interpreted as the one way distance
around a circle as shown in figure \ref{fcircle}. The metric is not
symmetric and the resulting graph is directed.
A triangle with sides length $i$ and $j$ with $i > j$ 
has the third side directed from the endpoint of $j$ and
is of length $d_{ji} = i-j$. Where the metric function is
now applied to lengths not node indices.
In the peer-to-peer context, this metric is used in the
Chord approach \cite{Chord}.

\bigskip
\noindent{\bf XOR metric.}
The length of a link between node labelled $a$ and $b$ is:
\begin{equation}
d_{ab} = a \oplus b
\end{equation}
Where $\oplus$ represents the bitwise XOR of the integer arguments.
In this case $N$ must be taken to be a power of 2 to
preserve the uniformity of the graph. It is possible to 
interpret this as the Manhattan distance when the nodes
are placed on a lattice as is shown for a simple case of 8 nodes
in figure \ref{fXOR}.
A consequence of this metric is that a triangle with sides 
length $i$ and $j$ has the third side length $d_{ij} = i \oplus j$.
In the peer-to-peer context, this metric is used in the
Kademlia approach \cite{Kademlia}.
\bigskip

The common structure of the graphs we consider arises from insisting that the
probability distribution function $p(d)$ that selects for the
existence of links of given distance is a strictly decreasing 
function of distance. 
At least for the case of the circle metric, this intuitively
gives rise to many short links along the perimeter of the
circle and fewer longer links as chords across the circle
along the lines of the small world proposal of 
Watts and Strogatz\cite{SmallWorld}.
We expect significant clustering as a consequence.
The picture is less clear for the XOR metric, but again
we anticipate some kind of small-world like structure to arise.
Such structure is desirable for a greedy routing algorithm as discussed in
the Introduction.

Preliminary investigations indicated that exponential decay 
of the probability distribution, even when appropriately scaled,
does not allow enough
long links for efficient routing and we have concentrated
effort on power law distributions:
\begin{equation}
p(d) = {z \over d^\alpha}
\end{equation}
The bulk of analysis will be for the case $\alpha=1$ but
to motivate this choice we retain it as a 
parameter for the present.
Were $\alpha$ to be
zero, the graph would be an
Erd\H{o}s and R\'enyi random graph with average degree $zN$,
and in general, $z$ has a related meaning 
that scales the average degree.
To bound the probability of length 1 links 
we take $0 < z \leq 1$, though it would be possible to consider
$z > 1$ provided short links are automatically present and
the probabilistic link selection only applies to longer links.
For example, with the circle metric, $z=1$ causes
all nearest neighbours to be connected and routability is guaranteed.   
This is not the case for the XOR metric which requires some
longer links in order to create a giant component.
In contrast to the case of Erd\H{o}s and R\'enyi random graphs,
$z$ does not scale with $N$ (as $1/N$ in that case), and consequently
the average degree depends on $N$.

\subsection{Graph Properties}

Before proceeding to consider routing let us briefly
characterise the $\alpha$ and $z$ parameter regimes according to some
standard graph properties.  We employ the techniques
described in \cite{GeneratingFunctions}
and use the following generating function for the
probability of vertex degrees.
\begin{equation}
G_0(x) = \prod_{i=1}^{N-1} \left(1 - {z(1-x)\over i^\alpha}\right)
\end{equation}
As follows from the uniform property of the graphs
without the need to specify which metric is used,
though in the case of the directed circle metric, this only counts
either the {\it in} or {\it out} degrees.

The moments of the degree distribution are computed
by taking derivatives of the generating function. For example
the first few central moments are:
\begin{eqnarray*}
\langle k\rangle &=& G_0'(1) 
= z \sum_{i=1}^{N-1} {1\over i^\alpha} = z H_{N-1,\alpha}\\
\langle (k-\langle k\rangle)^2\rangle &=& 
z H_{N-1,\alpha} - z^2H_{N-1,2\alpha}\\
\langle (k-\langle k\rangle)^3\rangle &=& 
z H_{N-1,\alpha} - 3z^2H_{N-1,2\alpha} + 2z^3H_{N-1,3\alpha}
\end{eqnarray*}
Where $H_{n,m}$ is the generalised Harmonic number
that in the limit
$n\to\infty$ becomes a Riemann Zeta function $\zeta(m)$.
Notice that for $\alpha < 1$ the average degree grows as $N^{1-\alpha}$. 
The shape of the degree distribution resembles that of a Poisson
law, but is narrower. 
Even for large graphs with $\alpha \le 1$ 
where some of the terms in the expressions above may be dropped,
the moments do not match those of a Poisson distribution.

The cluster coefficient \cite{SmallWorld} is defined as the ratio of the
number of triangles
to connected triples, $3N_\triangle/N_3$.
The number of connected triples can be
obtained from the vertex degree generating function as:
\begin{equation}
N_3 = {N\over 2}G_0''(1) =  
{N z^2\over 2} \left( (H_{N-1,\alpha})^2- H_{N-1,2\alpha} \right)
\end{equation}
Using the the uniqueness of lengths and the property of the metric
that determines the length of the third side of a triangle in terms of the
other two, it is straightforward to compute the probability of selecting 
links that form a triangle and consequently the number of triangles.
\begin{equation}
3N_\triangle = {Nz^3}
\sum_{i=2}^{N-1} \sum_{j=1}^{i-1} {1\over i^\alpha j^\alpha (d_{ij})^\alpha}
\end{equation}
The cluster coefficient is always proportional to $z$, but the sum above
is unwieldy for general  $\alpha$ and its value depends on whether the
graph is based on the circle or XOR metric.
For $\alpha > 1$ the coefficient has a finite limit for large $N$ graphs
but this is not generally the case for $\alpha \leq 1$.
At $\alpha = 1$ the sum becomes more tractable and for the XOR metric
we find:
\begin{equation}
{3N_\triangle\over N_3} =
{2z \over (H_{N-1})^2 - \zeta(2)} \sum_{i<j}^{N-1} {1\over i j (i\oplus j)}
\end{equation}
The unusual sum appearing in this formula is discussed in the appendix
and it approaches a finite limit of approximately  $1.54$ 
in the large $N$ limit.
Clustering vanishes slowly as $1/\log^2 N$ for large $N$ and 
simulations confirm the form derived above.

In the case of the circle metric more care is needed to take account
of the directed nature of the links. This gives rise to some
changes in factors but ends in the similar looking formula:
\begin{equation}
{3N_\triangle\over N_3} =
{3z \over 2(H_{N-1})^2 - \zeta(2)} \sum_{i<j}^{N-1} {1\over i j (j-i)}
\label{circletrianglecoeff}
\end{equation}
Here the limiting value of the sum can be expressed
as $2 \zeta(3)$, approximately 2*1.202057.

As a large connected component is a precondition for successfully
finding nodes, we want to ensure that the network is above any
possible percolation threshold. 
For $\alpha \le 1$ the average degree grows with $N$ and
in contrast to 
Erd\H{o}s and R\'enyi random graphs, we expect percolation
to occur for any value of $z$ \cite{RandomGraph}. 
This is indeed observed in
simulation but there is no simple proof since the
generating function techniques of \cite{GeneratingFunctions}
cannot be relied upon for this purpose as the existence of links is distance dependent.
We have just seen that for $\alpha > 1$
there is finite clustering in the thermodynamic limit and in this
range there may be a complex percolation threshold in the 
$\alpha$, $z$ plane. 
In the case of the circle metric at $\alpha \le 1$ a
rough test is to compute the probability that a gap 
with no links crossing it exists in the circle.
This probability goes to zero for any value of $z$, but
still there may be finite size effects at small $z$.

\medskip

In summary:
\begin{description}
\item[$\alpha > 1$,] 
$\langle k\rangle$ does not grow with $N$ 
and there is finite triangle clustering.
There may not be a giant component.
\item[$\alpha < 1$,] 
$\langle k\rangle$ is large and grows with $N$, 
but the triangle clustering coefficient decreases to zero.
\end{description}

As examples, we have simulated inverse square and inverse square root laws
and find the typical behaviour described above.  
Each regime has disadvantages. To be certain of
a giant component we should avoid  $\alpha > 1$,
but in the $\alpha < 1$ regime $\langle k \rangle$ grows.
To understand why this second issue is a problem   
we must start to consider routing issues. 
The average size of the table needed to keep track of neighbouring
nodes is the mean number of links leaving a node, 
that is, $\langle k \rangle$. 
For efficiency, the amount of information stored on a node 
should not grow too quickly with $N$. The parameter regime with 
$\alpha < 1$ is therefore less appealing.

The most interesting regime occurs at $\alpha = 1$
and for the rest of this paper we work at this point.
This leads to slow logarithmic growth of the routing table,
but potentially successful search. Moreover, this is 
precisely the situation motivated by Kademlia which
also has a table size that grows logarithmically.
Although we expect a giant component at $\alpha = 1$
and see one in simulations, we should beware of potential
finite size effects when the parameter $z$ is small.
Clustering does go to zero for large
graphs, but it does so as $1/\log^2 N$ which is much slower
than the $1/N$ expected for Erd\H{o}s and R\'enyi random graphs.
In fact, it is the clustering of nodes with nearby labels that is
important in this work rather than the global triangle cluster
coefficient, so it may still be legitimate to term 
these graphs small world.

\section{Greedy Routing}

 For the graphs we consider,
the only information the greedy routing algorithm requires at a node is the
list of nearest neighbours. This list contains the
node indices of the neighbours, and from this
information, the special distance metrics we have chosen
allow computation of the distance to that node and
moreover the remaining distance from the neighbouring
node to the final destination.

At each step the greedy algorithm chooses to hop to the neighbouring
node that is the closest to the eventual destination.
If there is no neighbour closer than the present node, 
then routing has failed according to this mechanism.
No backtracking is allowed. Note that the length
of the hop to the neighbour is not directly relevant to this
algorithm except though triangle bounds.

\begin{figure}
\epsfig{file=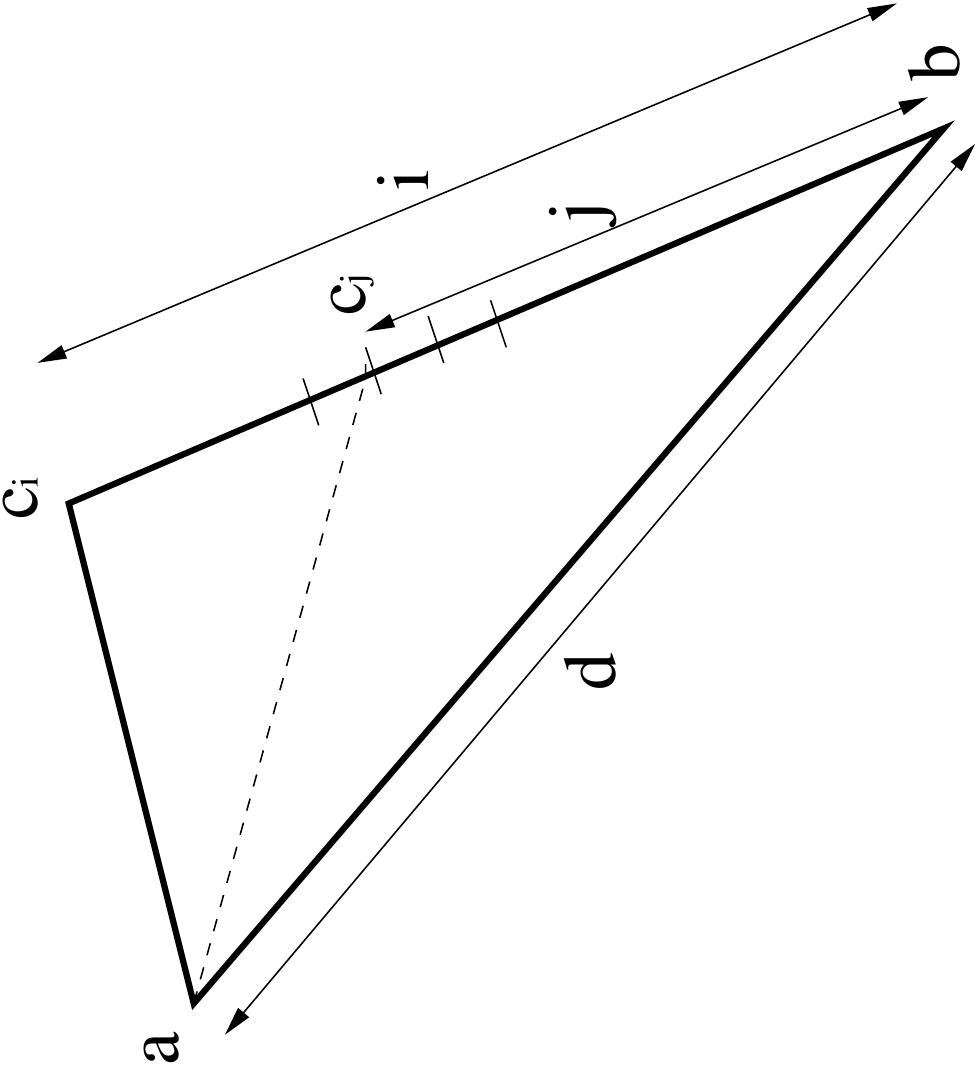,width=6cm,angle=270}
\caption{Routing from node $a$ to node $b$ via node $c_i$.
For the greedy algorithm to take this route 
there must be no link to any node $c_j$
closer to the destination than $c_i$.}

\label{frecurse}
\end{figure}

To analyse this algorithm consider the 
probability of successfully reaching the
destination at distance $d$ in $k$ hops, $q(d,k)$.
Of course the number of hops is limited by the distance $k\leq d$ as
each hop must get closer to the goal.
For the case of a single hop, 
$q(d,1)$ is merely the probability that a direct link exists.
\begin{equation}
q(d,1) = p(d)
\end{equation}
The probability of success in more than one hop can be
computed iteratively. Consider figure \ref{frecurse} 
for routing from node $a$ to node $b$ via node $c_i$. 
\begin{equation}
q(d,k+1) = \sum_{i=k}^{d-1}p(d_{ac_i}) q(i,k) 
\prod_{j=0}^{i-1} \left(1-p(d_{ac_j})\right)
\end{equation}
Where node labels and distances are as in the figure. 
The first terms are self evident and the product
accounts for the greediness of the algorithm by ensuring
that there is no neighbour $c_j$ of $a$
closer to $b$ than $c_i$.
Note that by virtue of the greediness of the algorithm,
$q(d,k)$ is independent of $N$ except that
it vanishes for $d > N$.

Using the triangle property of the metrics and
specialising to the $\alpha = 1$ probability distribution,
this can be written explicitly in terms of the metric function
for distances as:
\begin{equation}
q(d,k+1) = \sum_{i=k}^{d-1} {z\over d_{id}} q(i,k) 
\prod_{j=0}^{i-1} \left(1-{z\over d_{jd}} \right)
\end{equation}

To illuminate the routing algorithm and the recursion relation
it is helpful to consider $N=4$. We shall do this for the
XOR metric and the reader should have in mind an image of the graph
similar to that in figure \ref{fXOR}. The resulting probabilities are
shown in table \ref{tqofdkXOR}.
As an example of a two hop path consider $q(3,2)$. Starting
from node 0 there are two paths to the destination node 3: 
013 and 023 and in both cases
there is a factor $1-z/3$ to ensure that there is no direct path.
The preferred path would be 023 as the intermediate point is
closer to the destination: the contribution from this path
is $(z/2)q(1,1)(1-z/3)$. For the path 013, there is an additional
factor to exclude the possibility that the preferred path 023
exists and the contribution is $z q(2,1) (1-z/2)(1-z/3)$.
For the three hop case $q(3,3)$, notice that only one path
is possible: 0123 not 0213, as the step from 2 to 1 would move
further from the goal.
The $z$ dependence in the table agrees with numerical simulations.
For the  circle metric, the table is similar but the factors are
not identical.

\begin{table}
\begin{tabular}{|l|lll|}
\hline
$k$ & $d=1$ & $d=2$ & $d=3$\\ 
\hline
1 & $z$ & $z/2$ & $z/3$\\
2 & $0$ & $z^2(1-z/2)/3$ & $z^2(1-z/3)(1-z/4)$\\
3 & $0$ & $0$ & $z^3(1-z/2)^2(1-z/3)/3$\\
\hline
\end{tabular} 
\caption{Values of $q(d,k)$ for XOR metric for
small values of d}
\label{tqofdkXOR}
\end{table} 

It is convenient to define a generating function for the
probabilities $q(d,k)$.
\begin{equation}
Q(d,x) = \sum_{k=1}^{d} q(d,k) x^{k-1} 
\end{equation}
We shall use the recurrence relation that this obeys in analytic work,
but numerically it is more appropriate to directly consider 
quantities measurable in simulation.

By summing over all possible numbers of hops we obtain the 
overall probability $r(d)$ of routing over a distance $d$.
\begin{equation}
r(d) = \sum_{k=1}^{d} q(d,k) = Q(d,1)
\label{rofdsum}
\end{equation}
This quantity also obeys a recursion relation following 
from the relation for $q(d,k)$.
\begin{equation}
r(1) = z
\end{equation}
and
\begin{equation}
r(d) = {z\over d} + \sum_{i=1}^{d-1} {z\over d_{id}} r(i) 
\prod_{j=0}^{i-1} \left(1-{z\over d_{jd}} \right)
\end{equation}

In general $r(d)$ is a polynomial in $z$ of degree $d(d-1)/2$.
For very small $d$, the explicit form can be deduced by summing entries
from table \ref{tqofdkXOR}.


\section{Routability}

The recursion relations are complex, with each new value
depending on all previous ones. We therefore rely heavily on
numerical results, though we can throw some analytic light
on the system with the circle metric especially at $z=1$.
To numerically solve the recursion relations for
$r(d)$, values of $d$ up to $10^6$ are accessible 
in reasonable time on a desktop computer;
while for $q(d,k)$ we can only approach $d\sim 10^4$ with similar effort. 
We have also checked these
results and have investigated other properties by
directly simulating samples of the graphs and running the greedy 
routing algorithm on them. The simulations allow us to
investigate  properties such as the size of the
giant component, the triangle cluster coefficient
besides the probability of success and number of hops required 
for greedy routing. In reasonable time, graphs of 
sizes up to $32000$ can be studied with several hundred
samples. 

\subsection{Circle Metric}

For the circle metric the recursion relation for $r(d)$ becomes
\begin{eqnarray}
r(d) &= 
{z\over d} + \sum_{i=1}^{d-1} {z\over d-i} r(i) 
\prod_{j=0}^{i-1} \left(1-{z\over d-j} \right)\\
&=
{z\over d} + {z \Gamma(d+1-z)\over \Gamma(d+1)}
\sum_{i=1}^{d-1}  r(d-i) {\Gamma(i)\over \Gamma(i+1-z)}
\end{eqnarray}

\begin{figure}
\epsfxsize=0.5\hsize \epsfbox{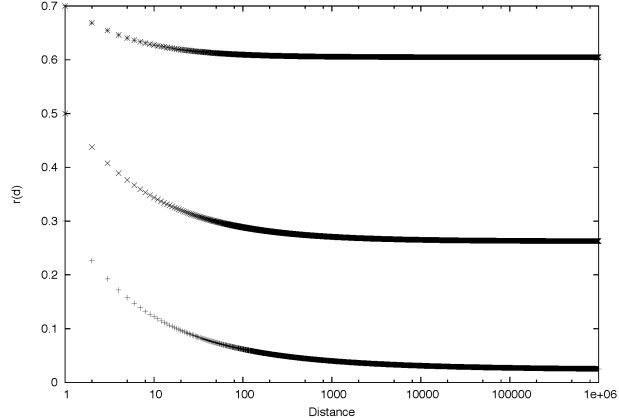}
\caption{The routing probability $r(d)$ on the circle
metric. From top to bottom at values of $z=$ $0.7, 0.5, 0.3$.}
\label{frofd}
\end{figure} 

Figure \ref{frofd} shows $r(d)$ for some representative values of $z$.
For $z=1$ routability is guaranteed since neighbours are
connected around the circle.
For values of $z$ in the approximate range $z>0.3$, 
the form $r(d) = r + a d^{-z}$
with $r$ and $a$ constant provides a very good fit
especially at larger $d$. Indeed, second order corrections
of the form $d^{-2z}$ can also be accurately identified.
This form is supported by 
analysis of the recursion relation. If constant
$r(d) = r$ is inserted on the right hand side, the
sum over gamma functions can be performed:
\begin{equation}
{z\over d} + {z \Gamma(d+1-z)\over \Gamma(d+1)}
\sum_{i=1}^{d-1}  r {z \Gamma(i)\over \Gamma(i+1-z)}
= r + {z(1-r)\over d} - r { \Gamma(d+1-z)\over \Gamma(d+1)\Gamma(1-z)}
\end{equation}
Note that part of the $1/d$ term is cancelled and that 
the remaining ratio of gamma functions
is indeed of order $d^{-z}$ at large $d$.
Unfortunately this still does not allow us to obtain an
analytic expression for $r$ because higher
order terms are needed that rely on
inserting the full (not asymptotic) form of
the corrections to $r(d)$ in the sum.
We shall return to a more careful asymptotic analysis below.

\begin{figure}
\epsfxsize=0.6\hsize \epsfbox{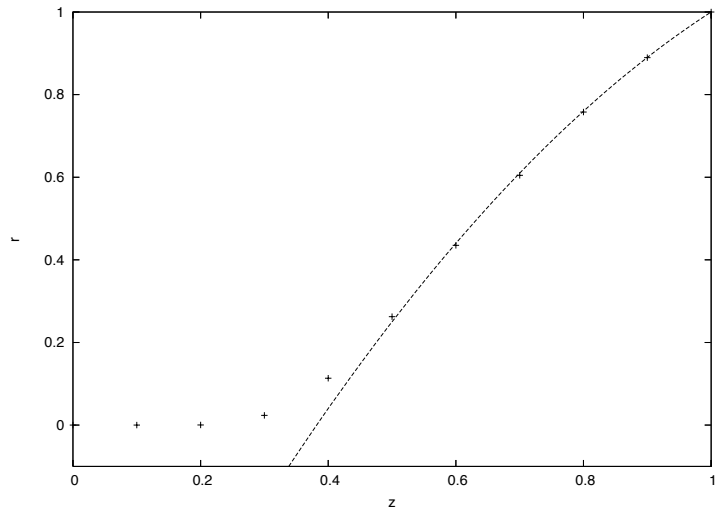}
\caption{The asymptotic constant value of $r(d)$ with the circle metric. 
Data points obtained by
numerically fitting to the form described in the text
using the last $5~10^5$ points of a solution of the
recursion relations to $d=10^6$. 
Error bars are too small to be visible.
Curve is the result of second order perturbation theory
about the point $z=1$.}
\label{frasyfit}
\end{figure} 

Simpler equations are obtained by 
perturbing away from the known behaviour at $z=0$ and $z=1$.
An expansion of $r(d)$ is inserted into the recursion relation
and although the resulting equations for the coefficients at each order
still depend on all coefficients at smaller $d$ as is the case 
for the full equation,
the structure is sufficiently simple to allow solution.

To second order  in $1-z$ we write:
\begin{equation}
r(d) = 1 - (1-z)r_1(d) - (1-z)^2 r_2(d)  
\end{equation}
Then the equation for $r_1(d)$ becomes:
\begin{equation}
r_1(d) = {1\over d} + {1\over d}\sum_{i=1}^{d-1} r_1(i)
\end{equation}
With solution $r_1(d)=1$ independent of $d$.
The equation for the second order coefficient is:
\begin{equation}
r_2(d) = {H_{d-1}\over d} + {1\over d}\sum_{i=1}^{d-1} r_2(i)
= {1\over d}\sum_{i=1}^{d-1} \left(r_2(i)+ {1\over i}\right)
\end{equation}
This has the solution $r_2(d) = 1 -1/d$.
The finite size correction $d^{-z}$ seen numerically 
and included in the asymptotic analysis is
expected to appear as a log at third order.

Combining these terms and taking the 
large $d$ limit to find the asymptotic constant part of $r(d)$:
\begin{equation}
r = 1 -(1-z) - (1-z)^2
\end{equation}
In figure \ref{frasyfit}
this curve is shown alongside the 
numerically determined value of 
the asymptotic constant $r$
and is a good match for larger values of $z$
The value of $r$ vanishes below
a critical value of $z_c = (3-\sqrt 5)/2 = 0.38197$. 
This is in the correct vicinity of a numerical transition, and
provides support for the existence of this transition.

The equations for the perturbative expansion around $z=0$ have simpler
structure, no longer involving all coefficients at smaller values of $d$,
but the terms appearing in the equations have more
complicated analytic form.
For small values of $z$ the perturbative series 
to second order is:
\begin{equation}
r(d) = {1\over d}z + {2 H_{d-1}\over d} z^2
\end{equation}
Where each term separately vanishes in the large $d$ limit. However, the
higher order terms which are expected to be of the form $(\log d)^{n-1} z^n/d$,
do so slowly
and examples below will provide a warning that this kind
of series can easily sum to a constant.

The numerical data in  figure \ref{frasyfit}
is only able to indicate that
the asymptotic constant $r$ becomes very small for $z$
below the putative transition. We would like to 
investigate the asymptotic properties of the
equations more closely in order to gain more evidence
for a transition.
As a model for this we start
by considering the special value $z=1$ 
where the equations simplify to the extent that 
progress can be made. 
It is straightforward to check that $r(d)=1$ is a solution
of the recursion relations without any need to take limits.
But to study this in more detail we look at the probabilities
$q(d,k)$ which include the hop information. On the circle the
recursion relation becomes:
\begin{eqnarray}
q(d,k+1) &=& {z \Gamma(d+1-z)\over \Gamma(d+1)}
\sum_{i=k}^{d-1}  q(i,k) {\Gamma(d-i)\over \Gamma(d-i+1-z)}\nonumber\\
 &\stackrel{z=1}{=}& {1\over d}
\sum_{i=k}^{d-1}  q(i,k)
\label{qofdkZ1} 
\end{eqnarray}
And we can obtain the first few values immediately:
\begin{eqnarray}
q(d,1) &=& {1\over d}\nonumber \\ 
q(d,2) &=& {1\over d} H_{d-1}\nonumber\\ 
q(d,3) &=& {1\over 2 d} \left( (H_{d-1})^2 - H_{d-1,2}\right)\nonumber\\ 
q(d,4) &=& {1\over 3! d}\left( (H_{d-1})^3 - 3H_{d-1}H_{d-1,2} + 2H_{d-1,3}\right)
\label{qcircle}
\end{eqnarray}
Where the $H_{n,m}$ are generalised Harmonic numbers.

For a general solution it is better to return to the form:
\begin{equation}
q(d,k) = {1\over d}
\sum^{d-1}_{i_{k-1} > i_{k-2}\cdots>i_2 > i_1 = 1} {1\over \prod_{j=1}^{k-1} i_j}
\end{equation}
Then by an exercise in combinatorics \cite{KnuthV1} 
the generating function for the $q(d,k)$ is:
\begin{eqnarray}
Q(d,x) &=& \sum_{k=1}^{d} q(d,k) x^{k-1}\nonumber\\
&=& {1\over d}\prod_{i=1}^{d-1} \left( 1 + {x\over i}\right)\nonumber\\
&=& {1\over d}\exp\left(-\sum_{k=1}^\infty {H_{d-1,k} (-x)^k\over k}\right)\nonumber\\
&\stackrel{d\to\infty}{=}&
{1\over d^{1-x} \Gamma(1+x)}
\label{Qzequals1}
\end{eqnarray}
The generating function obeys the relation $(d+1)Q(d+1,x)=(d+x)Q(d,x)$.
By expanding, the general form of $q(d,k)$ along the lines of
(\ref{qcircle}) can be written as
a sum over partitions. A recurrence relation for $q(d,k)$ in terms of the
generalised Harmonic numbers also follows.

To recover the $z=1$ result that $r(d) = 1$ 
by summing $q(d,k)$ according to formula (\ref{rofdsum}),
some care is needed. Although the correct result is obtained 
by simply summing the leading asymptotic term 
${\log^{k-1}d/d(k-1)!}$, there is no reason not to expect 
subleading terms to also contribute.
We proceed using the generating function:
\begin{eqnarray}
r(d) &=& Q(d,1)\nonumber\\
&=& {1\over d}\exp\left(H_{d-1}-\sum_{k=2}^\infty {H_{d-1,k} (-1)^k\over k}\right)\nonumber\\
&\stackrel{d\to\infty}{=}&
{1\over d}\exp\left(\gamma + \log d -\sum_{k=2}^\infty {\zeta(k) (-1)^k\over k}\right)\nonumber\\
&=& 1
\end{eqnarray}
Where the result follows at finite $d$ from interchanging the order
of sums on the second line, but we have proceeded to the limit
using an identity relating the alternating sum of zeta
functions to the Euler Mascheroni constant $\gamma$.

This approach, still at $z=1$, can be extended to compute the 
expectation value for the number of hops.
\begin{eqnarray}
\langle k \rangle 
&=& {1\over r(d)}\sum_{k=1}^d k q(d,k)\nonumber\\
&=& 1 + Q'(d,1)\nonumber\\
&=& 1  - \sum_{k=1}^\infty H_{d-1,k}(-1)^k\nonumber\\
&=& H_d\nonumber\\
&\stackrel{d\to\infty}{=}& \gamma + \log d
\label{meankzequals1}
\end{eqnarray}
This form is confirmed numerically and acts as a check of the
numerical implementation.

Using intuition gained from the solution of the $z=1$ case, 
we wish to perform a similar analysis for $z<1$.
The analysis is based on the
recurrence for the generating function:
\begin{equation}
Q(d,x) = 
{z\over d} + {xz \Gamma(d+1-z)\over \Gamma(d+1)}
\sum_{i=1}^{d-1}  {\Gamma(d-i)\over \Gamma(d-i+1-z)} Q(i,x)
\label{Qcirclerecur}
\end{equation}
We search for an asymptotic solution
resembling the last line of (\ref{Qzequals1}) of the form:
\begin{equation}
Q(d,x) = 
{c_1\over d^{1-\beta}}
+{c_2\over d^{1-\beta+z}} + \dots
\label{Qanzatz}
\end{equation}
Where the coefficients $c_i$ and the exponent $\beta$ are functions
of $z$ and $x$. We insert this expansion on the right hand side of
(\ref{Qcirclerecur}) relying on the fact that a similar procedure
works at $z=1$. Then using Euler Maclaurin to estimate the sum in terms
of an integral and matching powers of $d$ we find an equation
for the exponent, a series of equations relating coefficients,
and from the $1/d$ term,
a normalisation equation for the coefficients that involves
them all. Only the equation for the exponent can be solved in isolation:
\begin{equation}
{\Gamma(\beta + z)\over \Gamma(\beta)}
= x \Gamma(1+z)
\end{equation}
This correctly reproduces $\beta = x$ at $z=1$ and from
graphical considerations it is clear that there is a unique
solution $\beta(x,z)$ for all values of $z$ and $x$ in their
range. At $x=1$ we find $\beta=1$ for all values of $z$, so
the leading term in $r(d) = Q(d,1)$ is constant and the
exponent alone is unable to indicate a transition. 
Fortunately, the expectation value for the 
mean number of hops does provide a way
of accessing the value of $\beta$, or at least its derivative:
\begin{equation}
\langle k \rangle = 1+ {Q'(d,1)\over Q(d,1)}
= 1 + {c_1'\over c_1} + \beta' \log d
\end{equation}
Where the derivatives are with respect to $x$ and 
all terms are evaluated at $x=1$.
The coefficient of the log is given by:
\begin{equation}
\beta' = {1\over \psi(1+z) +\gamma}
\label{betaprime}
\end{equation}
Where $\psi$ is the dilogarithm.


\begin{figure}[htbp]
  \centering
  \begin{minipage}[b]{5 cm}
\epsfxsize=1.0\hsize \epsfbox{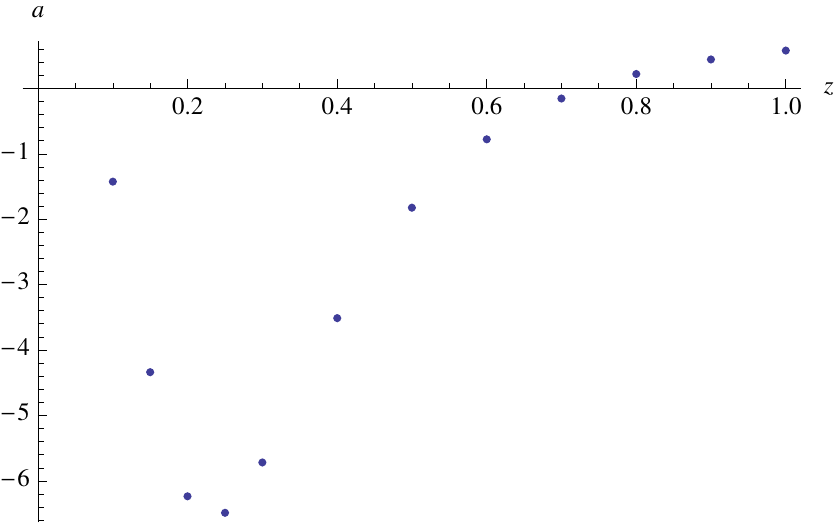}
  \end{minipage}
\qquad
  \begin{minipage}[b]{5 cm}
\epsfxsize=1.0\hsize \epsfbox{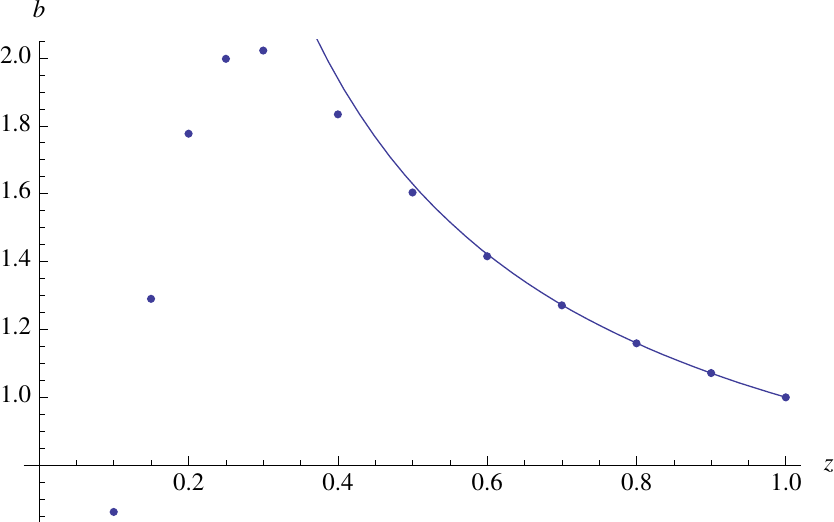}
  \end{minipage}
\caption{The coefficients $a$ (left) and $b$ (right) of a logarithmic 
fit $\langle k \rangle = a + b \log d$ 
using the last $5000$ points of a numerical solution 
on the circle metric to $d=10^4$. 
Error bars are too small to be visible.
The curve in the figure for $b$ arises from the asymptotic analysis
discussed in the text.}
\label{avkcoeffabfig}
\end{figure}

Numerically,  a logarithmic form is a good fit
to the expected number of hops and indeed  at $z=1$,
(\ref{meankzequals1}) is exact and was used as a check of the
computer implementation of the recursion relation for $q(d,k)$.
For all $z<1$ the numerical curves of $\langle k\rangle$ 
are accurately fitted by $a + b \log(d)$ and the
coefficients $a$ and $b$ are shown as functions of
$z$ in figure \ref{avkcoeffabfig}.
The prediction of (\ref{betaprime})
is shown in the second of these figures
and is accurate for values of $z$ above the transition.
Below that point, the
prediction continues to grow, but the data reverses its
trend. This is presumably a finite size effect as in
the thermodynamic limit the probability of successful
routing vanishes below the transition 
and $\langle k \rangle$ cannot be defined.

Returning to the asymptotic analysis of 
(\ref{Qcirclerecur}),
the normalisation equations for $c_i$ involve all
the coefficients and we cannot solve them to obtain any expression for 
$r = c_1$. However, the ratio $c_2/c_1$ of the sub-leading to
leading coefficients can be computed.
Again this yields favourable comparison with numerical
results in the region above the transition. 
However, in neither of the cases where it has been
tested, has the asymptotic analysis indicated the
existence of a transition. Based on the clear
disagreement with the numerical results and the
evident difficulty in separating the leading and subleading
terms in (\ref{Qanzatz}) for small $z$
we conclude that our asymptotic analysis fails in this region
and is unable to provide information about the transition.


\begin{figure}
\epsfxsize=0.5\hsize \epsfbox{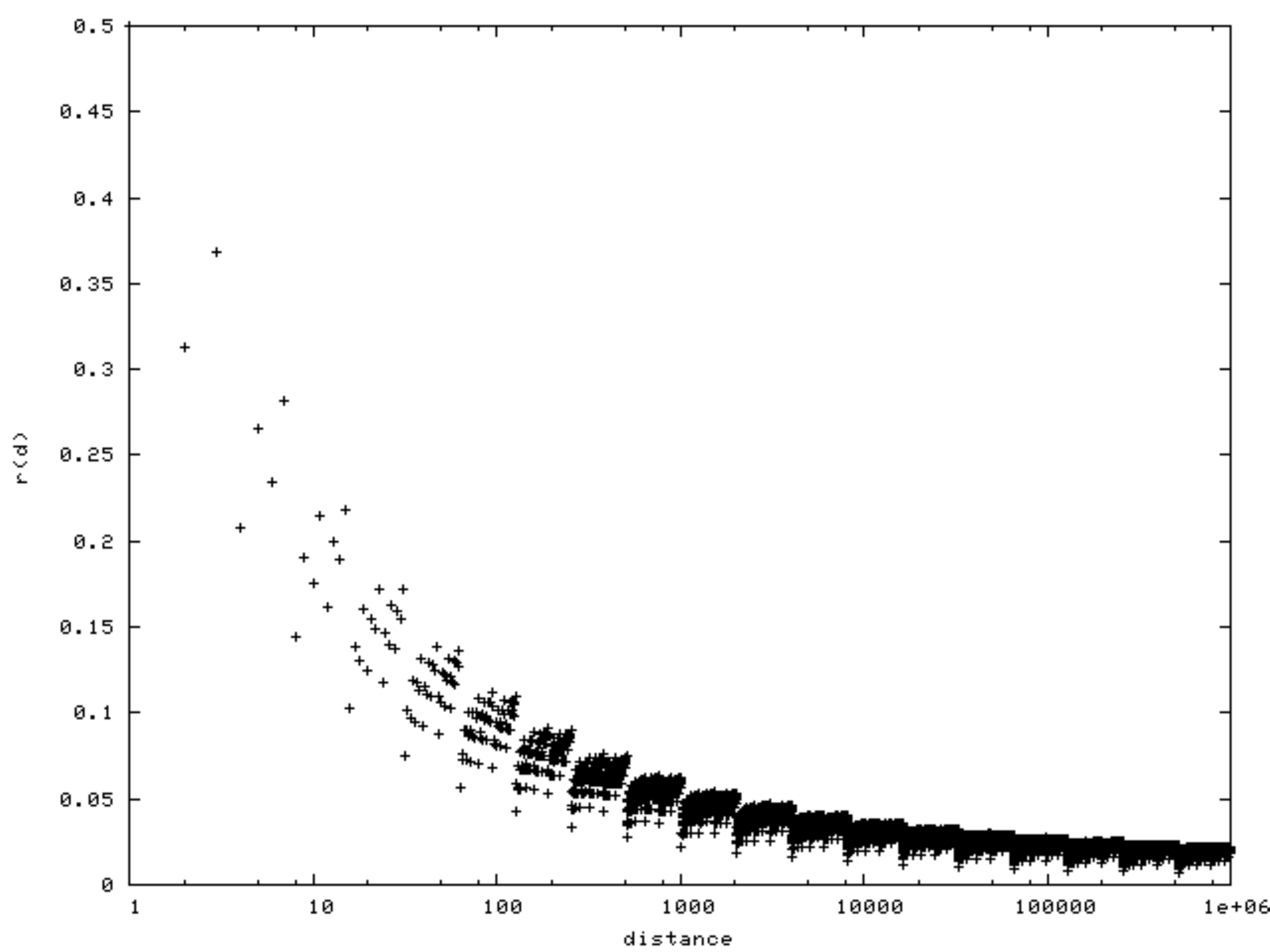}
\caption{The routing probability $r(d)$ on the XOR
metric at $z=0.5$.}
\label{frofdx}
\end{figure}

\subsection{XOR Metric}

For the XOR metric the recursion relations are less
amenable to analytic work even at the special value $z=1$. 
Numerical solution
of the following recurrence for $r(d)$ is shown in figure \ref{frofdx}.
\begin{equation}
r(d) = 
{z\over d} + \sum_{i=1}^{d-1} {z\over d \oplus i} r(i) 
\prod_{j=0}^{i-1} \left(1-{z\over d \oplus j} \right)
\end{equation}
The large fluctuations in $r(d)$ as $d$ increments
by small amounts are characteristic of all results with this metric 
and it is 
therefore not appropriate to fit the whole curve.
We have selected points at $d=2^m$ and at $d=2^m-1$ which
appear to bound $r(d)$ from below and above respectively and
provide smooth curves that can be fitted. 
The fitting procedure 
itself differs from that used on the circle, not least
because there is no reason to expect the exponent to 
take the value $z$. 
In order to avoid numerical instability, we have chosen
to fit to the form $a^2 + b^2/d^c$ where positivity of the coefficients
is enforced and the exponent $c$ is also fitted. 
Since the selected data points are spaced exponentially 
the choice of which points to use in the fit is also
different, and based on runs up to $d=10^6$ 
all points above $d=1000$ have been chosen. 

\begin{figure}[htbp]
  \centering
  \begin{minipage}[b]{5 cm}
\epsfxsize=1.0\hsize \epsfbox{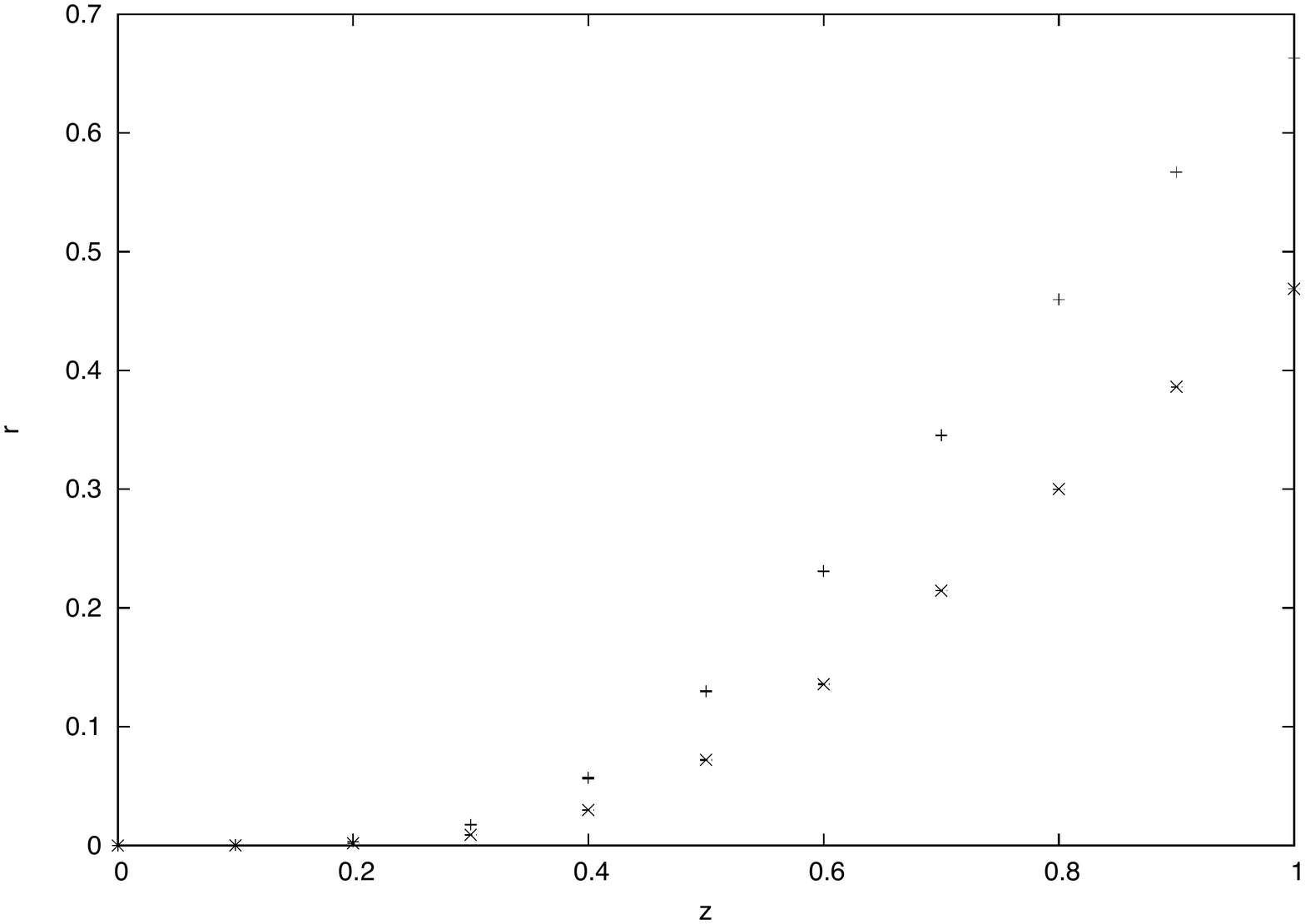}
  \end{minipage}
\qquad
  \begin{minipage}[b]{5 cm}
\epsfxsize=1.0\hsize \epsfbox{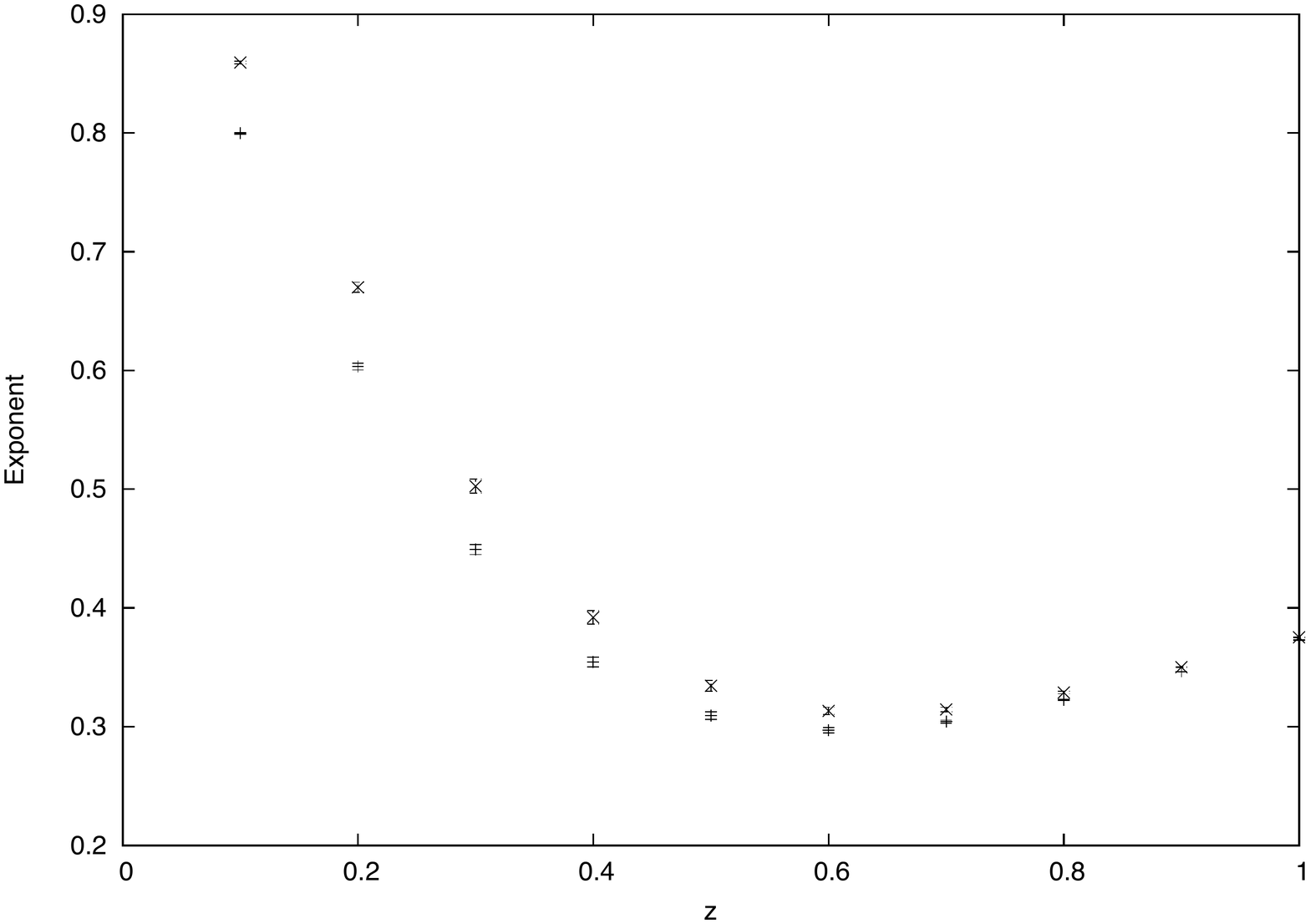}
  \end{minipage}
\caption{Results of a numerical fit to $r(d)$
based on the fitting procedure described in the text.
The left hand plot shows the asymptotic constant part $r=a^2$
and the right hand plot shows the exponent
of the finite size corrections.
On the left hand the upper points are a fit for $d=2^m-1$ and the
lower points are for $d=2^m$ while this identification is
reversed for the right hand plot.
}
\label{rasyfitfigx}
\end{figure}

The results of these fits for the constant part and the exponent are
shown in figure \ref{rasyfitfigx}.
Notice that the routing success is below one even for $z=1$ and that
the exponent certainly does not behave as $z$
so the situation is rather different from that for the circle metric. 
The exponent takes a similar value for both the bounds, 
and we might anticipate that 
there is a common exponent to describe all values of $d$.
The asymptotic constant $r=a^2$ takes distinct values for each
bound but both curves appear to converge to indicate a transition in the
same vicinity.

\begin{figure}[htbp]
  \centering
  \begin{minipage}[b]{5 cm}
\epsfxsize=1.0\hsize \epsfbox{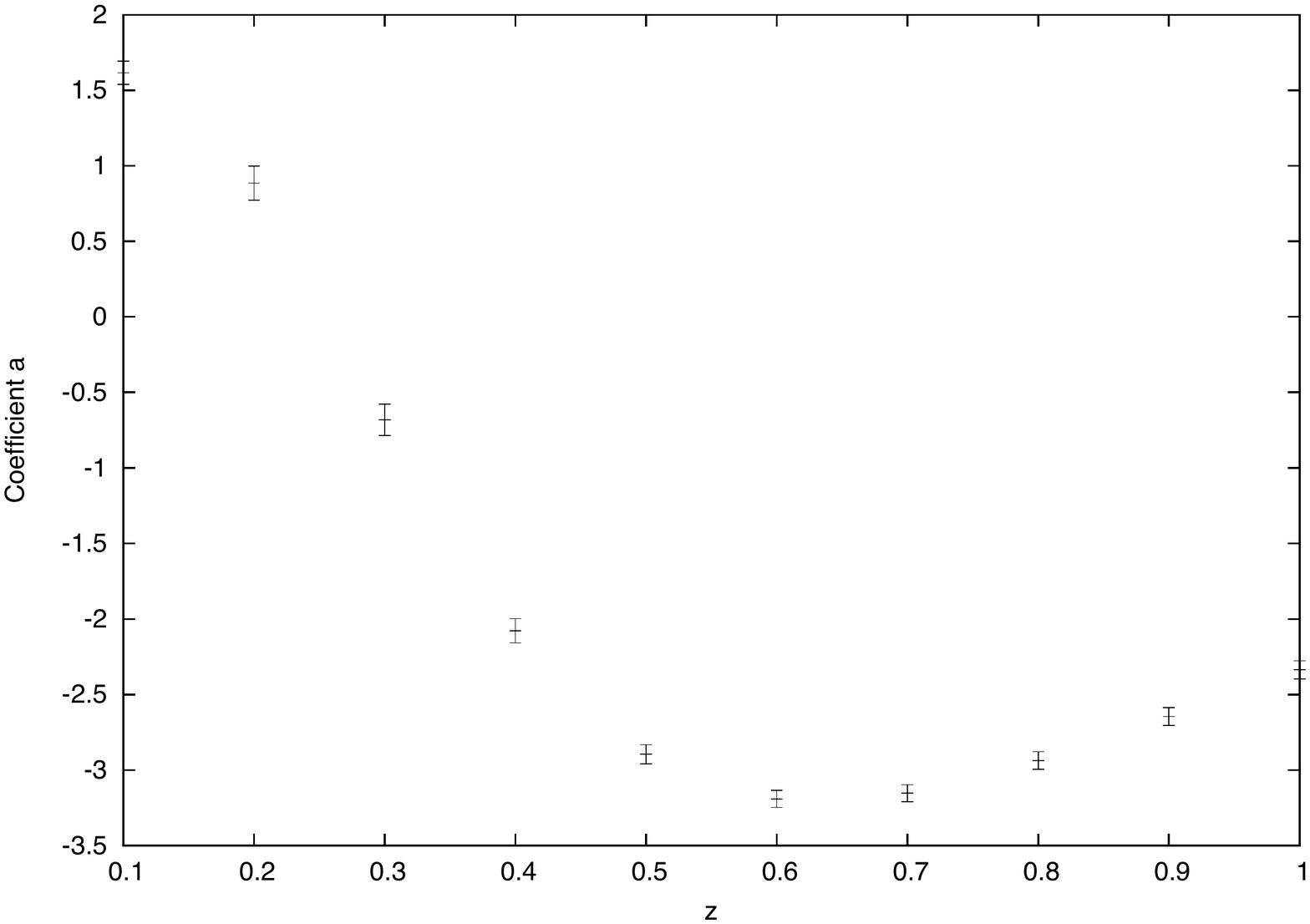}
  \end{minipage}
\quad
  \begin{minipage}[b]{5 cm}
\epsfxsize=1.0\hsize \epsfbox{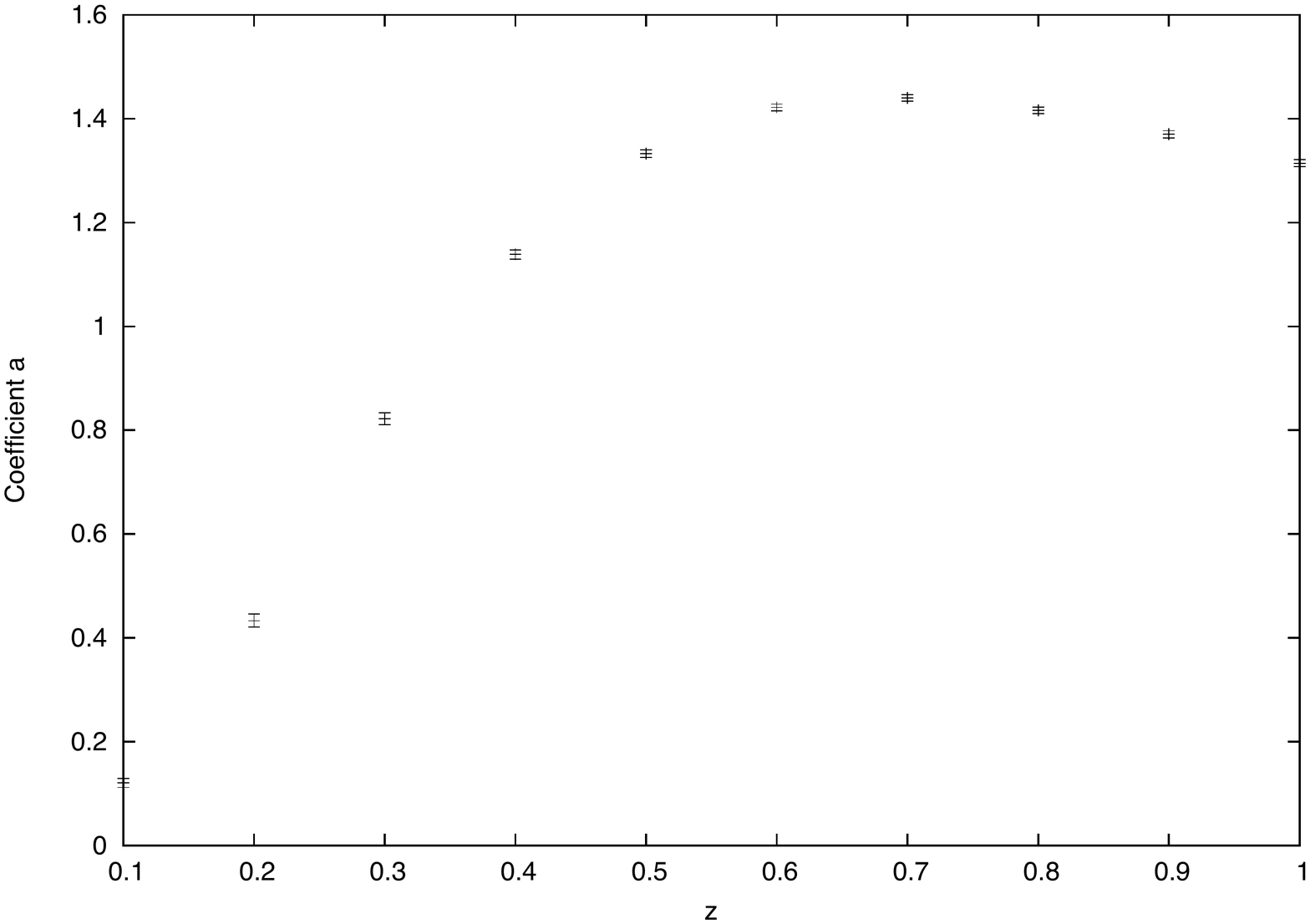}
  \end{minipage}
\caption{The coefficients $a$ (left) and $b$ (right)
of a logarithmic fit to $\langle k \rangle$ as obtained by
numerically fitting to the form described in the text
using the last $5000$ points of a run to $d=10^4$. 
Error bars are shown.}
\label{avkcoeffabxfig}
\end{figure}

Although the XOR metric leads to substantially more fluctuation in
the mean number of hops $\langle k \rangle$ than the circle metric, the
overall trend remains logarithmic. 
In the two figures \ref{avkcoeffabxfig}
the coefficients $a$ and $b$ of a fit to the form 
$a + b \log(d)$ are shown.
These curves are less abrupt than for the circle metric
but follow the same trend. Notice in particular that
the average number of hops required by the XOR metric
is less than that needed by the circle metric when
searching for nodes an equal distance away and
moreover that it retains a much more stable value in
the region above the transition.



\section{Conclusion}

We have considered the problem of efficiently finding a 
labelled node on a random graph using a simple greedy
algorithm that makes
decisions of which link to take solely on the basis 
of the list of labels of neighbouring nodes to which it is directly connected.
To enable this, the random graph was
constrained to have small world like structure, 
and we proposed and constructed a class of graphs
with properties intended to
facilitate the search.

In the limit of very large size
graphs we have demonstrated that for an appropriate range
of parameters, the search for a node arbitrarily far away
has a finite probability of success. 
Moreover, we have strong numerical hints that the system 
displays a transition between the regime with finite probability
of locating a desired node at large $z$ to a regime where
this is not guaranteed at smaller $z$. 
Unfortunately, the  $\log N$ dependence of many quantities 
makes it hard for numerical work to accurately predict behaviour in the
thermodynamic limit.
The existence of the transition for the circle metric 
has some support from analytic
analysis, with both the perturbative and asymptotic analysis 
matching numerical results at larger $z$, and with the
perturbative approach indeed predicting a transition at 
finite $z$.  For the XOR metric where analytic work is much harder,
the graphs obtained from numerical solution of the equations
have similar shape to those for the circle metric, and we
expect the same conclusion.

In our work
it seems that the random graphs based on XOR metric are 
less likely to lead to successful routing than those based
on the circle. The reason for this observation is that 
while the sum of the hop distances is equal to the total
distance to the end node on the circle, for the XOR metric
it is greater, and consequently the probability of such
links existing is smaller. On the other hand, the
number of hops needed to reach the destination is typically
fewer for XOR than circle metric.
However, Kademlia has generally been favoured
over Chord in existing present day peer-to-peer systems.
This merely emphasises the difference in approach to our work as
peer-to-peer systems are dynamic and the list of stored
links changes in order to optimise
routing which is eventually always successful. 
The XOR metric is symmetric so the results of any query can be
used to update the local table, thus minimising the number
of queries needed. 


\vskip 0,5 truecm

\noindent{\bf Acknowledgment:} 
I would like to thank
M.Z.~Ahmed and  B.~Ghita for conversations.

\appendix
\section{Sums involving XOR}
Here we examine some unusual sums that
involve the XOR operator.
The precise value of these sums fluctuates
within an envelope, so it is the bounds that are studied.

First it is helpful to investigate.
\begin{equation}
S_1(n) = \sum_{i=1}^{n-1} {1\over i (i\oplus n)}
\end{equation}

For $n=2^m-1$ and for the range of $i$ appearing in the sum,
it becomes apparent that $n\oplus i = n-i$
by considering the binary form of $n$.
 In this case, $S_1 = 2 H_{n-1}/n$
and this constitutes the lower bound.

For $n=2^m$ similar considerations lead to 
$n\oplus i = n+i$ and in this case,
$S_1 = 2H_{n-1}/n - H_{2n-1}/n +1/n^2$ which acts as
an upper bound.

By taking the $n\to\infty$ limit we find that
\begin{equation}
{\log n\over n} < S_1(n) < {2\log n\over n}
\label{s1bounds}
\end{equation}

\bigskip
Now consider the more elaborate sum that appears
in the expression for the triangle clustering coefficient.
\begin{equation}
S_2(n) = 
\sum_{i=2}^{n-1} \sum_{j=1}^{i-1} {1\over i j (i\oplus j)}
=\sum_{i=2}^{n-1} {S_1(i)\over i}
\end{equation}
Note that this obeys the recurrence relation:
\begin{equation}
S_2(n+1) = S_2(n) + {S_1(n)\over n}
\end{equation}
This allows us to determine the form of the finite
size corrections to the asymptotic constant.
\begin{equation}
S_2(n) \to const - {b \log n\over n}
\end{equation}
The bounds on $S_1(n)$ in (\ref{s1bounds}) translate into bounds on the
parameter $b$, but the main purpose of this exercise is to
justify the form of corrections and thus allow accurate
numerical determination of the asymptotic constant.
This is necessary because in contrast to the circle case
where the recurrence relation can be used
the full sum must be performed in numerical work. 
This restricts the sizes accessible.

For the circle metric, the sum in equation (\ref{circletrianglecoeff})
has the same type of corrections to the asymptotic form, but does
not suffer from the variation indued by XOR.


 
\pagestyle{plain}

\end{document}